\documentclass[sw]{agutex}
\usepackage{rotating}
\usepackage{float}
\usepackage{graphicx}
\setkeys{Gin}{draft=false}
\authorrunninghead{BLAKE, ET AL.}

\titlerunninghead{GICs in the Irish Power Network}

\authoraddr{Corresponding author: Se\'{a}n P. Blake,
School of Physics, Trinity College Dublin, Dublin 2, Ireland
(blakese@tcd.ie)}

\begin{document}

%% ------------------------------------------------------------------------ %%
%
%  TITLE
%
%% ------------------------------------------------------------------------ %%

\title{Geomagnetically Induced Currents in the Irish Power Network during Geomagnetic Storms}
%
% e.g., \title{Terrestrial ring current:
% Origin, formation, and decay $\alpha\beta\Gamma\Delta$}
%

%% ------------------------------------------------------------------------ %%
%
%  AUTHORS AND AFFILIATIONS
%
%% ------------------------------------------------------------------------ %%

%Use \author{\altaffilmark{}} and \altaffiltext{}

% \altaffilmark will produce footnote;
% matching \altaffiltext will appear at bottom of page.

 \authors{Se\'{a}n P. Blake,\altaffilmark{1,2}
 Peter T. Gallagher,\altaffilmark{1} Joe McCauley,\altaffilmark{1}
 Alan G. Jones\altaffilmark{2, 5},
 Colin Hogg\altaffilmark{2},
 Joan Campany\`{a}\altaffilmark{2},
 Ciar\'{a}n D. Beggan\altaffilmark{3},
 Alan W.P. Thomson\altaffilmark{3},
 Gemma S. Kelly\altaffilmark{3},
 David Bell\altaffilmark{4}}

\altaffiltext{1}{School of Physics, Trinity College Dublin, Dublin 2, Ireland}
\altaffiltext{2}{Dublin Institute for Advanced Studies, 5 Merrion Square, Dublin 2, Ireland}
\altaffiltext{3}{British Geological Survey, Lyell Centre, Riccarton, Edinburgh, EH14 4AP, UK}
\altaffiltext{4}{EirGrid Plc., The Oval, 160 Shelbourne Rd, Ballsbridge, Dublin 4, Ireland}
\altaffiltext{5}{Complete MT Solutions, 5345 McLean Crescent, Manotick, Ontario, K4M 1E3, Canada}

%\altaffiltext{2}{Department of Geography, Ohio State University,
%Columbus, Ohio, USA.}

%\altaffiltext{3}{Department of Space Sciences, University of
%Michigan, Ann Arbor, Michigan, USA.}

%\altaffiltext{4}{Division of Hydrologic Sciences, Desert Research
%Institute, Reno, Nevada, USA.}

%\altaffiltext{5}{Dipartimento di Idraulica, Trasporti ed
%Infrastrutture Civili, Politecnico di Torino, Turin, Italy.}

%% ------------------------------------------------------------------------ %%
%
%  KEYPOINTS
%
%% ------------------------------------------------------------------------ %%

% Key points are 1 to 3 points that the author provides,
% that are 100 characters or less, that are ultimately published
% with the article.
%% for example:
% \keypoints{\item Here is the first keypoint. what happens if it is a
% long keypoint, like this one. We want to see this wrap please.
% \item This is the second.
% \item And here is the third keypoint
% }

\keypoints{\item Surface electric fields and geomagnetically induced currents (GIC) were simulated in the Irish power network for five geomagnetic events.
\item A multi-layered resistivity model to a depth of 200~km was made using magnetotelluric measurements for use in GIC simulations in Ireland.
\item GICs have been replicated for Kp6 and Kp7 storms, and predicted for Kp9 storms in Ireland.}

%% Keypoints will print underneath the abstract.

%% ------------------------------------------------------------------------ %%
%
%  ABSTRACT
%
%% ------------------------------------------------------------------------ %%

% >> Do NOT include any \begin...\end commands within
% >> the body of the abstract.

\begin{abstract}
Geomagnetically induced currents (GICs) are a well-known terrestrial space weather hazard. They occur in power transmission networks and are known to have adverse effects in both high and mid-latitude countries. Here, we study GICs in the Irish power transmission network (geomagnetic latitude 54.7--58.5$^{\circ}$ N) during five geomagnetic storms (06-07 March 2016, 20-21 December 2015, 17-18 March 2015, 29-31 October 2003 and 13-14 March 1989). We simulate electric fields using a plane wave method together with two ground resistivity models, one of which is derived from magnetotelluric measurements (MT model). We then calculate GICs in the 220, 275 and 400~kV transmission network. During the largest of the storm periods studied, the peak electric field was calculated to be as large as 3.8~V~km\textsuperscript{-1}, with associated GICs of up to 23~A using our MT model. Using our homogenous resistivity model, those peak values were 1.46~V~km\textsuperscript{-1} and 25.8~A. We find that three 400 and 275~kV substations are the most likely locations for the Irish transformers to experience large GICs. \textbf{Accepted for publication in AGU Space Weather. Copyright 2016 American Geophysical Union. Further reproduction or electronic distribution is not permitted. DOI: 10.1002/2016sw001534}

\noindent 
%respectively.

\end{abstract}

%% ------------------------------------------------------------------------ %%
%
%  BEGIN ARTICLE
%
%% ------------------------------------------------------------------------ %%

% The body of the article must start with a \begin{article} command
%
% \end{article} must follow the references section, before the figures
%  and tables.

\begin{article}

%% ------------------------------------------------------------------------ %%
%
%  TEXT
%
%% ------------------------------------------------------------------------ %%

\section{Introduction}

Geomagnetic induced currents (GICs) are the most hazardous phenomena associated
with space weather. They manifest most prominently during intense geomagnetic storms
as quasi-DC currents that flow through man-made conductors such
as gas pipelines (e.g., \citet{Campbell1986, Pulkkinen2001}) and electric power transmission grids (see review papers \citet{Viljanen1994} and \citet{Pirjola2000}). In extreme geomagnetic storms, they have the potential to disrupt transmission systems, as happened in Canada in the March 1989 storm \citep{Bolduc2002}. Extreme GICs can directly damage transformers through spot-heating \citep{Zheng2013}. This damage to a transformer can be costly and can potentially leave areas without power for extended periods of time as well as incurring the cost of replacing the transformer. As such, both the physical and economic effects of geomagnetic induced currents have been widely studied in many countries (e.g., \citet{Piccinelli2014, Schrijver2014}). 

It has been recognised that high geomagnetic latitudes (greater than $60^\circ$)
are at particular risk from GICs where geomagnetic disturbances are
larger and more frequent \citep{Pirjola2000}. GICs have therefore been
studied extensively in northerly regions, particularly Scandinavia
\citep{Pulkkinen2001, Wik2008, Myllys2014}. While less at risk, countries with latitudes less than $60^\circ$ have also been found to experience GICs in their power networks. These include the UK \citep{Beamish2002, Beggan2013}, 
New Zealand \citep{Marshall2012}, Spain \citep{Torta2014}, China \citep{Zhang2015} and South Africa \citep{Ngwira2011}.
It is now known that one way that GICs can contribute to the failure of transformers in low and mid-latitude
countries is through repeated heating of the transformer insulation \citep{Gaunt2007, Gaunt2014}. As such, 
countries which previously disregarded GICs in their networks may in fact have had transformer damage
due to space weather effects and misattributed the cause of the damage.

While the March 1989 storm is perhaps the most famous example of GICs causing damage to power infrastructure, it is not the largest storm on record. The 1859 ``Carrington Event'' storm has been estimated to have been approximately 1.5 times as geoeffective as the 1989 storm \citep{Siscoe2006} and occurred at a time when countries were not reliant on power networks. If a storm of this magnitude were to happen today, it would likely cause widespread GICs in power networks across the world. In July 2012, Earth experienced a ``near-miss'' with a  powerful coronal mass ejection (CME) \citep{Baker2013}. This particular CME would have given rise to a geomagnetic storm larger than even the 1859 event had it been Earthward directed. These kind of events show that large storms may be rare, but can happen at any part of the solar cycle. It is therefore important to study the response of power networks in countries where GICs might not typically be considered a large risk.

The most straightforward approach to studying GICs in a power network is to measure them at transformers. This can be done directly using a Hall effect probe to measure the current flowing in transformer earth connections during periods of geomagnetic activity \citep{Thomson2005, Torta2014}. GICs can also be measured by utilising the differential magnetometer method, where magnetometers are placed near power lines to measure the magnetic signal of DC GICs \citep{Matandirotya2016}. Where a Hall effect probe is not available and no direct measurements of GICs exist, GICs can be modelled in two steps. The first step is to calculate the surface electric fields across the region of interest. Estimates for the electric field can be calculated by combining geomagnetic data with a conductivity or resistivity model. Studies have utilised region-wide estimates for ground conductivities \citep{Adam2012, Wei2013}. Methods for calculating geoelectric fields  range from the simple plane-wave method \citep{Pirjola1985, Viljanen2004} to more complicated methods such as a thin-sheet approximation \citep{McKay2003, Thomson2005} or complex image method \citep{Pirjola1998}. The second step to calculating GICs is to apply the calculated electric fields to a model power network. GICs can be treated as quasi-DC ($<$1 Hz) currents that flow through transformer grounded neutrals \citep{Pirjola2001}.

Ireland has no recorded instances of transformer damage which has been attributed to geomagnetic activity. Ireland also had no Hall effect probes to directly measure GICs until September 2015, when a probe was installed on a 400~kV transformer in the east of the country. As such, GIC measurements during geomagnetic storms are limited. GICs can be calculated in the manner described above in order to investigate the effect historical storms may have had on the Irish power grid. 

In this paper, we present the first detailed study of GICs in the Irish power grid for multiple geomagnetic storms. Geomagnetic data from INTERMAGNET observatories, as well as Irish observatories which make up the new Irish Magnetometer network (MagIE), were interpolated across Ireland using the spherical elementary current system method (SECS) for different geomagnetic storms. Surface electric fields were then calculated using the plane-wave method along with a magnetotelluric (MT) derived earth resistivity model. Finally, a model of the Irish 400~kV, 275~kV and 220~kV transmission grid was combined with the calculated electric fields and resultant GICs were calculated.

In this manner, two recent local K6 storms (20--21 December 2015 and 06--07 March 2016) were simulated, and their resultant GICs compared to GIC measurements at a transformer near Dublin in the east of the country. More severe historical geomagnetic storms were then simulated in the same way. These were the 13--14 March 1989, 29--31 October 2003 and 17-18 March 2015 storms.

\section{Observations}

\subsection{Geomagnetic Field Measurements} \label{FieldMeasurements}
The five geomagnetic events studied in this paper can be separated into two categories. These are the recent events (20--21 December 2015 and 06--07 March 2016) and the historical events (13--14 March 1989, 29--31 October 2003, 17--18 March 2015). The recent events were chosen as they are the largest geomagnetic storm events to give unambiguous GIC measurements at the Hall effect probe on the transformer near Dublin since its installation in September 2015. The three historical events were chosen as they are well studied examples of large events (registering as 8-, 9o and 9o on the planetary K-index respectively) with good geomagnetic coverage.

\begin{center}
\begin{figure}
\noindent\includegraphics[width=20pc]{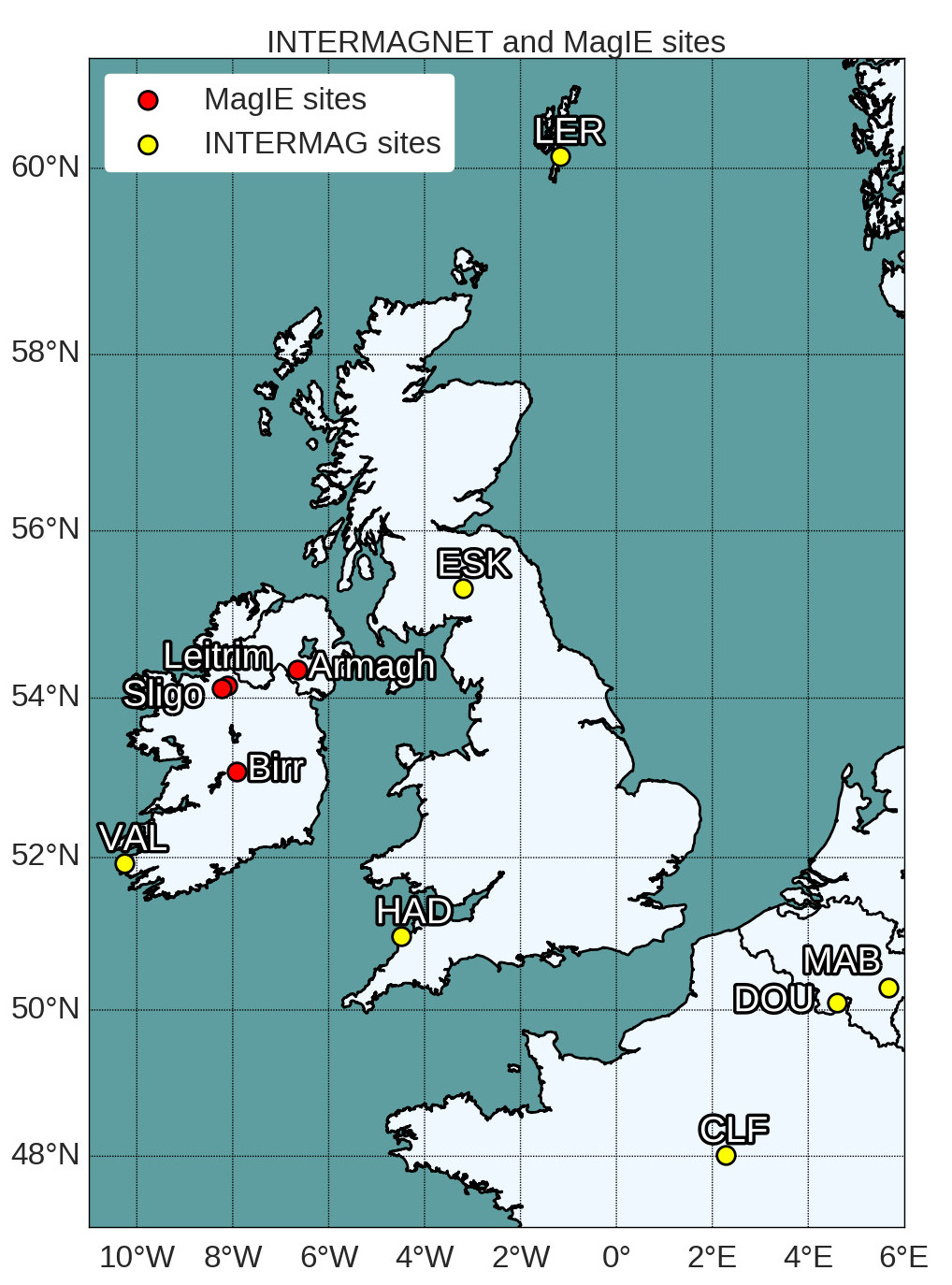}
\caption{Location of INTERMAGNET and MagIE sites in Ireland, Britain and continental Europe. The Leitrim site was moved 10~km west to Sligo in late 2015. Armagh is functinoning as part of the MagIE network, but was not operational during any of the events studied in this paper. The location of the MagIE sites are: Birr 53.09$^\circ$ N, 7.92$^\circ$ W;
Sligo 54.12$^\circ$ N, 8.22$^\circ$ W; Leitrim 54.16$^\circ$ N, 7.92$^\circ$ W and Armagh 54.35$^\circ$ N, 6.65$^\circ$ W.}
\label{mag_sites}
\end{figure}
\end{center}

Geomagnetic observations were taken from a variety of different magnetometer stations located in Ireland, Britain and continental Europe. 
The stations located in Ireland are Valentia (VAL), Birr, Sligo and Leitrim. The stations located in Britain are
Hartland (HAD), Eskdelamuir (ESK) and Lerwick (LER). The continental stations are 
Chambon la For\^{e}t (CLF), Manhay (MAB) and Dourbes (DOU). The locations of these observatories are shown in Figure \ref{mag_sites}.

Data from these sites were taken from different magnetometer observatory networks, depending on availability. The first of these is the MagIE, a network of magnetometers and electrometers which measure changes in both local geomagnetic and geoelectric fields due to space weather, and has been in operation
since late 2012.
One-second geomagnetic time series from Birr in central Ireland are currently available at the Rosse Solar-Terrestrial
Observatory website (www.rosseobservatory.ie). Data from other MagIE sites 
will be available online from late 2016 at the same website.

The International Real-Time Magnetic Observatory Network (INTERMAGNET; www.intermagnet.org) hosts 1-
minute geomagnetic
data dating from 1991 to the present. INTERMAGNET and MagIE observations were used for
the two recent storm events, 20--21 December 2015 and 06--07 March 2016, as well as the 17--18 March 2015
event. Only INTERMAGNET observations
were used for the 29--31 October 2003 event. As INTERMAGNET does not host data prior to 
1991, the World Data Centre for Geomagnetism, Edinburgh (www.wdc.bgs.ac.uk) was used
for the stations which were active during the March 1989 storm. An exception to this is Valentia, whose 
data were supplied by Met \'{E}ireann.

All data, where necessary, were averaged into 1-minute bins. Any gaps in the time-series
were estimated using a linear interpolation. No gaps in the magnetic data were greater than
20 minutes. An average baseline for each event was subtracted for each storm period studied. The 
stations used for each event can be found in Table \ref{stations_table}.

\begin{table*}
\caption{List of stations used for each of the geomagnetic events studied in this paper.}
\begin{center}
\begin{tabular}{| l | r |}
\hline 
Event Date & Sites Used \\
\hline  \hline
06-07 March 2016 & Birr, Sligo, VAL, ESK, HAD, LER \\
20-21 December 2015 & Birr, Sligo, ESK, HAD, LER \\
17-18 March 2015 &  Leitrim, Birr, VAL, ESK, HAD, LER, CLF, DOU, MAB \\   
29-31 October 2003 & VAL, ESK, HAD, LER, CLF, DOU \\ 
13-14 March 1989 &  VAL, ESK, HAD, LER, CLF \\
\hline
\end{tabular}
\end{center}
\label{stations_table}
\end{table*}

\subsection{Ground Resistivity Model}
For the calculation of GICs, three different resistivity models were used. The first of these 
is the \citet{Adam2012} Europe-wide model. For Ireland, this consists of three different regions with 
varying resistivity values down to 30~km. All values deeper than this are set at 200~$\Omega$~m in the 
model. 

The second model used is a simple homogenous Earth with a resistivity of 100~$\Omega$~m. That is to say, 100~$\Omega$~m across Ireland and at all depths.

The third model is an multi-layered resistivity structure with values derived from over 750 individual 
MT sites. From 2004--2014, the Dublin Institute for Advanced Studies (DIAS) conducted 
a number of different geophysical projects around Ireland to map the conductivity of the Irish 
lithosphere at various depths using MT data. These projects were conducted with a 
number of different scientific objectives in mind, from crustal geometry definition to geothermal 
energy to carbon sequestration. A byproduct of these different MT 
surveys is that the measurements taken can be used when calculating electric-fields for studying GICs.

\begin{center}
\begin{figure*}
\includegraphics[width=30pc]{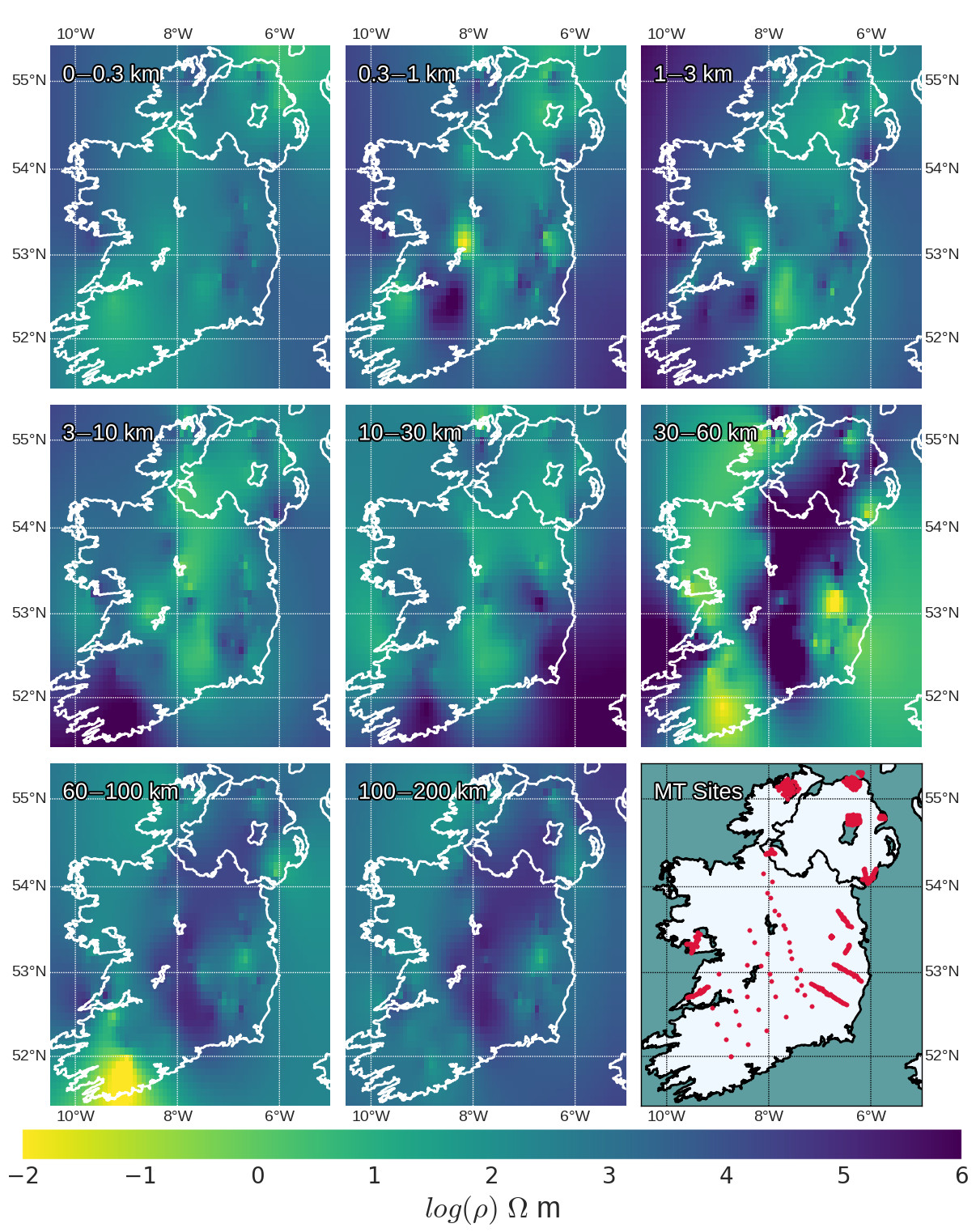}
\caption{Resistivity of various depth intervals in the Irish geology down to 200~km as given by our MT model. The bottom right plot shows the location of the sites from different MT surveys which informed the model. The values given by the MT sites were interpolated across Ireland using a radial basis function for different depths.}
\label{conductivity_map}
\end{figure*}
\end{center}

Each of the MT sites used in the model gave an average resistivity for different depths of the Irish 
subsurface down to 200~km. The values from these sites were interpolated using a linear radial basis 
function onto a 10~km $\times$ 10~km grid (shown in Figure \ref{conductivity_map}). These 
values were then used to make a layered Earth model with values for 0--0.3~km, 0.3--1~km, 1--3~km, 
3--10~km, 10--30~km, 30--60~km, 60--100~km and 100--200~km. A resistivity value of 100~$\Omega$~m was set for depths greater than 200~km. 
This is a value for Ireland which was chosen as it best fit our GIC observations. The nature of the MT 
projects undertaken by DIAS means that a majority of the points are confined to a few dense regions of 
geological interest: i.e., in areas where there are sandstone basins. The remainder of the points 
originate from larger region-wide surveys, such as in \citet{Rao2014}.

The spatial distribution of the sites means that particularly the west and southwest of Ireland have 
large areas where there are no MT measurements to bound the interpolation scheme. This could 
potentially skew resistivity values and therefore electric field calculations for those regions. The error which could arise from these skewed values are offset however by the method of surface electric field calculation used in this paper. These methods require that only the surface electric fields directly beneath transmission lines are known. As all of the substation nodes in the transmission system model used in this paper are within 40~km of one of the MT sites, this limits the error that the spatial distribution introduces when calculating GICs.

\subsection{Power Transmission Network}

\begin{center}
\begin{figure}
\includegraphics[width=21pc]{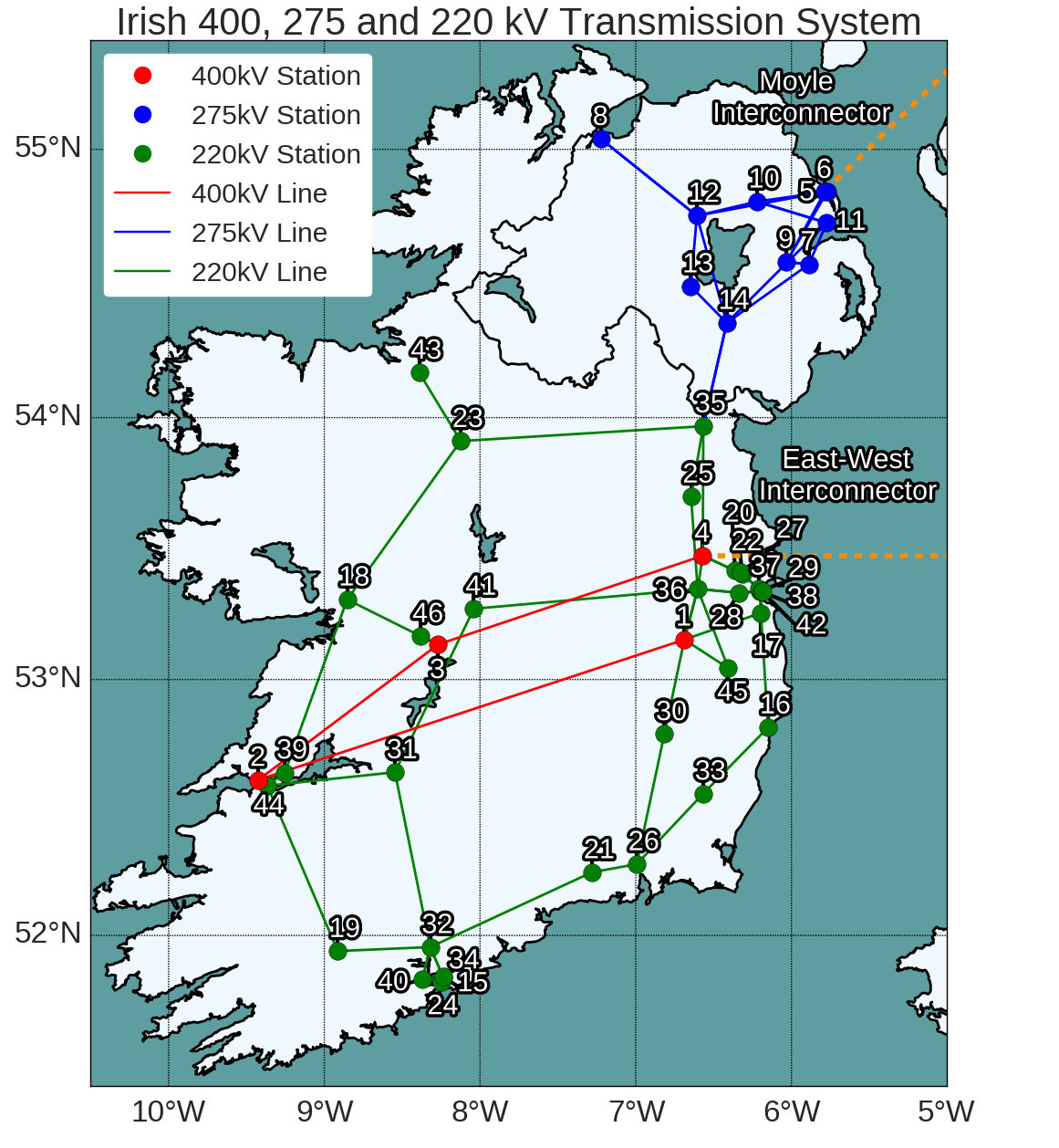}
\caption{The 400, 275 and 220 kV power transmission system in Ireland. Substation numbers are plotted beside each substation. Each of the nodes were connected with straight line paths, although true distances were used for calculating transmission line resistances. Substations are ordered by voltage (400~kV: 1--4, 275~kV: 5--14, 220~kV: 15--46) and then alphabetically by the name of each substation. Although not included in the simulations in this paper, the dashed orange lines represent the two HVDC interconnectors which link the Irish power grid to that of Britain. Substation 4 is where Ireland's only GIC probe is installed.}
\label{power_network}
\end{figure}
\end{center}

The power network in Ireland as of 2016 is composed of 400, 275,
220 and 110~kV transformers and transmission lines. For this study,
the lower voltage 110 kV transformers and shorter 110~kV lines were omitted. The
make-up of the grid is such that the Northern Ireland transmission system has 275~kV lines and 
transformers, whereas the Republic of Ireland has 400~kV and 220~kV 
lines and transformers. The 220~kV lines roughly follow the coastlines, whereas the three 400 kV lines 
run in a roughly north-easterly direction through the centre of Ireland. 
For simplicity, transformer nodes were connected with straight transmission
lines, and line resistances were calculated from line composition and true length. The longest transmission line measures 209~km.
Grounding resistances for all transformers were assumed to be 0.1 $\Omega$, with transformer resistances assumed to be 0.5~$\Omega$. These are approximate values which are used frequently where true resistances are not known \citep{Myllys2014, Torta2014}. 
General details of the model grid are given in Table \ref{grid_stats}, and the model grid is shown in 
Figure \ref{power_network}.

\begin{table}
\caption{General details of the model grid used in the simulations.}
\begin{center}
\begin{tabular}{| l | r |}
\hline \hline
Average Length of Connection &  41 km \\   
Max Length of Connection & 209 km \\   
Average Line Resistance & 2.13 $\Omega$ \\ 
Number of Nodes & 46 \\
Number of Transmission Lines & 78 \\
\hline
\hline
\end{tabular}
\end{center}
\label{grid_stats}
\end{table}

\section{Geoelectric Field and GIC Modelling}

For each event, the modelled GICs were calculated in two steps:
\begin{enumerate}
\item The geoelectric field was calculated for every 10~km $\times$
10~km square using an interpolated magnetic field and the conductivity model as inputs for 
the plane-wave method. The magnetic field required for this was calculated for each minute interval 
using the spherical elementary current system method.

\item The model transmission network was imposed onto the calculated electric
field and GICs were calculated at each network node.
\end{enumerate}

Further details of the above method are given in the following sections.  All GICs in this paper are expressed as neutral currents.

\subsection{Geomagnetic Field Modelling}

The varying horizontal magnetic field was calculated for every 10
km $\times$ 10 km square by utilizing the SECS method \citep{Amm1999}. 
This method interpolates the horizontal magnetic field at a given location from known measurements.
It achieves this by assuming that the magnetic field on the ground can be 
represented by a system of divergence-free equivalent currents in the ionosphere. As such, it neglects 
entirely any internal geomagnetic field component (i.e., any geomagnetic field which is a result of the 
Earth's internal structure or subsurface conductivity). These ionospheric currents are solved for the 
known magnetic fields (i.e., magnetometer stations), and can then be used to calculate the unknown 
magnetic field at a different location.

SECS has been shown to reproduce the varying magnetic field for large sparse arrays on a continental 
scale \citep{McLay2010}. However, the area in which the magnetic fields are being replicated in 
this study is approximately 300 $\times$ 500 km, and with the combination of the MagIE and
INTERMAGNET observatories, Ireland is well covered by true magnetic readings for SECS. For more 
detailed and localised studies, more magnetometer stations are required to enhance the spatial density of measurements.

As mentioned above in Section \ref{FieldMeasurements}, different magnetic observatories were operating for each of the five geomagnetic
storms studied (see Table \ref{stations_table}). Of the five events, the March 1989 storm has the poorest magnetic coverage,
with only five of the chosen magnetic observatories recording in Ireland, Britain and mainland Europe.
To test the efficacy of the SECS method for this case, the measured geomagnetic data at the Birr
observatory in central Ireland for the 17-18 March 2015 event were compared with SECS modelled magnetic
data. The modelled data were calculated using only the geomagnetic observatories which were operational
during the 1989 storm. Figure \ref{SECS_Validation_1989} shows the measured and modelled $B_x$ 
and $B_y$ components for the event, as well as the horizontal magnetic components for each of the sites 
used. A common evaluation of how well a model fits measurements is to calculate the root mean square 
difference ($\mathrm{RMSD}$). This is calculated as follows:

\begin{center}
\begin{figure}
\includegraphics[width=20pc]{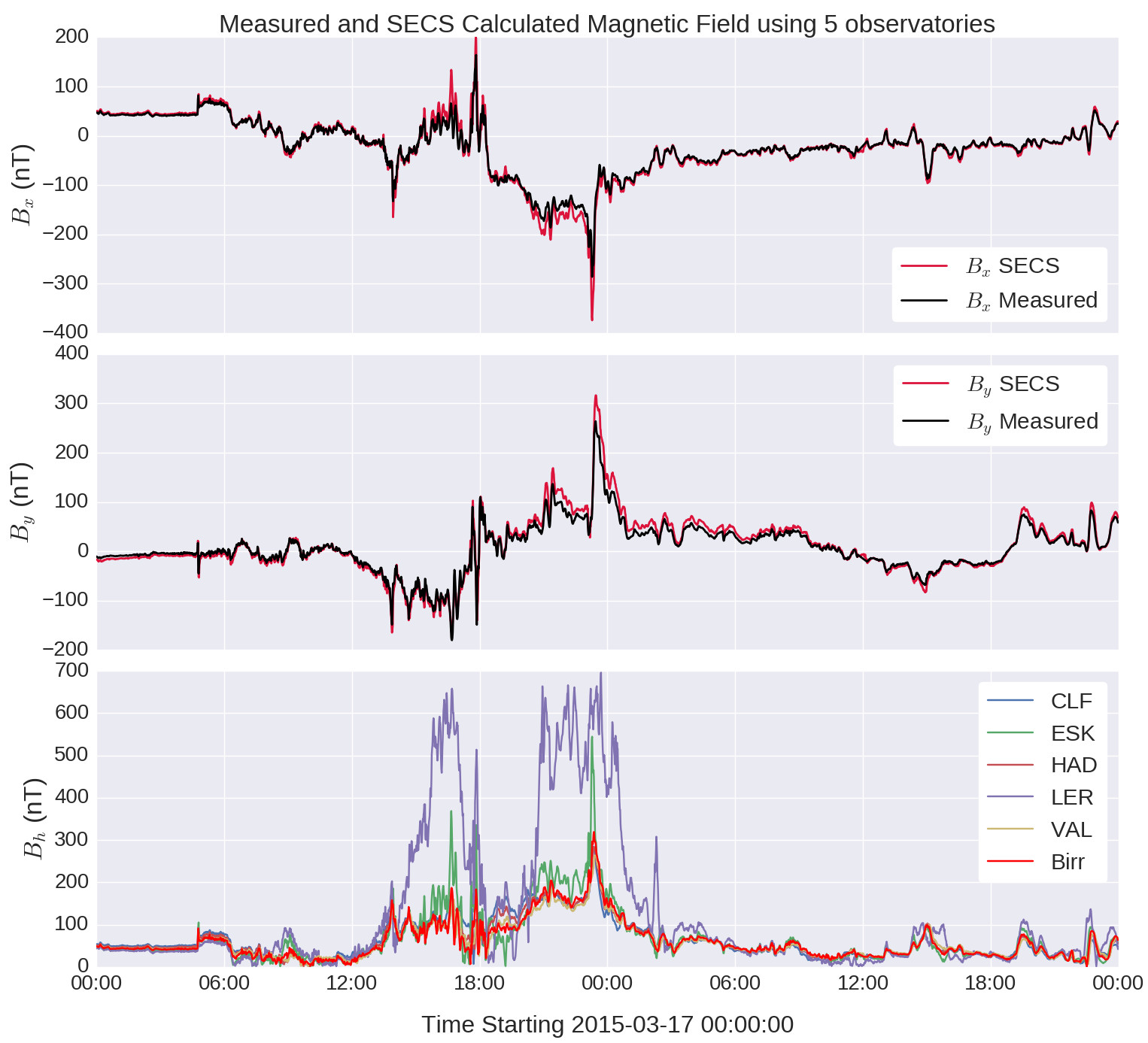}
\caption{Top and middle: The measured and SECS calculated $B_{x}$ and $B_{y}$ magnetic field components at Birr for the 17-18 March 2015 storm. The magnetic field was calculated using only the observatories which were available during the 13-14 March 1989 storm (VAL, LER, ESK, HAD and CLF). The horizontal component from these sites, along with Birr,  are shown in the bottom plot. The RMSD for the estimated and measured $B_{x}$ and $B_{y}$ components at Birr are 8.8 and 10.5 nT respectively.}
\label{SECS_Validation_1989}
\end{figure}
\end{center}

\begin{equation} \label{eq:1}
\mathrm{RMSD_{oc}} =\sqrt{\frac{\Sigma_{i=1}^{N}(o_{i}-c_{i})^{2}}{N}}
\end{equation}
where $o_{i}$ and $c_{i}$ are the $i$th observed and calculated points from a total of $N$. 
The $\mathrm{RMSD}$ error for the $B_x$ and $B_y$ components in this instance are 8.8~nT and 10.5~nT respectively for the two day period. This RMSD value increases to 20.2 and 20.3 for the most variable part of the storm (13:00 UT on the 17th to 01:00 UT on the 18th of March).

It is worth noting that although this RMSD is small when compared to the absolute values measured, the error for the interpolated data scales with intensity. At the peak of the storm, the magnetic field was overestimated by about 30\% in the $B_x$ component. This is due to the lack of true magnetic sites in Ireland north of Valentia. The addition of the MagIE sites mitigate this problem for the later events.

\subsection{Geoelectric Field Modelling}

\noindent 
The plane wave method is the simplest way of relating surface electric and magnetic fields widely in use for calculating GICs (see \citet{Pirjola2001} for details of this and other methods). Its core equation assumes a plane electromagnetic wave which propagates down into a layered or uniform Earth. The
frequency-dependent ($\omega$) plane-wave equation which describes the relation between horizontal electric and magnetic field components at a surface is given by 

\begin{equation} \label{eq:2.1}
\mathbf{E}(\omega) = \mathbf{Z}(\omega)\mathbf{B}(\omega)
% \end{equation}
\end{equation}

or

\begin{equation} \label{eq:2.2}
\left( \begin{array}{cc}
E_{x}\\
E_{y}\\
\end{array} \right) = \frac{1}{\mu_{0}}
\left( \begin{array}{cc}
Z_{xx} & Z_{xy}\\
Z_{yx} & Z_{yy}\\
\end{array} \right)
\left( \begin{array}{cc}
B_{x}\\
B_{y}\\
\end{array} \right)
\end{equation}

\noindent where $\mathbf{Z}$ is the magnetotelluric or impedance tensor, and $\mu_{0}$ is the vacuum permeability \citep{Chave2012}. $\mathbf{Z}$ is dependent on resistivity structure, and is calculated by iteratively relating the impedance at the top and bottom of each layer of the Earth \citep{Cagniard1953}.
For a 1-D Earth resistivity structure (i.e., where the resistivity changes only with depth, and not laterally), the impedance tensor becomes

\begin{equation}
\mathbf{Z}_{1D} = \left( \begin{array}{cc}
0 & Z_{xy}\\
-Z_{xy} & 0\\
\end{array} \right)
\end{equation}

\noindent This sets the parallel elements ($Z_{xx}$ and $Z_{yy}$) to zero as lateral changes are ignored, leaving only the off-diagonal elements $Z_{xy}$. The electric field components can then be written as 

\begin{equation}
E_{x}(\omega) = \frac{1}{\mu_0} Z_{xy}(\omega) B_{y}(\omega)
\end{equation}

and

\begin{equation}
E_{y}(\omega) = \frac{-1}{\mu_0} Z_{xy}(\omega) B_{x}(\omega)
\end{equation}

\noindent These are the frequency dependent equations which were used in this paper to calculate the electric field when 
coupled with the MT derived multi-layered resistivity model.

If we use a uniform ground resistivity model, by inverse-Fourier transforming Equation 4, we obtain a time domain relation between the electric and magnetic fields \citep{Torta2014}.

\begin{equation}
E_{x,y}(t)=\pm\frac{1}{\sqrt{\pi\mu_{0}\sigma}}\int_{0}^{\infty}\frac{1}{\sqrt{\tau}}\frac{dB_{y,x}(t-\tau)}{dt}d\tau
\end{equation}

\noindent where $dB/dt$ is the varying magnetic field component perpendicular to $E$, $\tau$ is a time increment 
and $\sigma$ is a single conductivity value. This equation is the plane wave approximation assuming a 
uniform Earth. This is equivalent to equation 4 if $Z_{xy}$ was calculated for an Earth with a single 
resistivity value. Equation 5 was discretised according to \citet{Pirjola1985} for the purposes of this 
paper.

As the plane-wave method is ultimately a simplification, it does not take into account spatial changes 
in conductivity, i.e., coastal effects. This is in contrast to more complicated methods of calculating 
electric fields such as the thin-sheet approximation \citep{Thomson2005, Vasseur1977}.

% Horizontal geoelectric field data were measured for both 2015 storm events and the March 2016 event in 
% MagIE's Leitrim and Sligo sites (the Leitrim site was moved $\sim$10~km West in late 2015 to Sligo). 
% Both horizontal measured and calculated E-field components in Leitrim for the 17-18 March 2015 storm 
% are shown in \textbf{Figure} \ref{E_verification}. Both the MT and homogenous Earth approaches 
% considerably overestimate the electric-field. This is likely due to the nature of the local ground
% where our electrodes are installed. In the case of Leitrim, this is a generally wet and boggy cut 
% forest.

\subsection{Geomagnetic Induced Current Modelling}
The DC approach as specified by \citet{Viljanen1994} was used to calculate GICs in this study. The 
currents flowing to and from substations in this study are expressed as the sum of current through each 
phase. A summary of the method is outlined below.

A power transmission system can be represented as a discrete system with \textbf{N} earthed nodes
(transformer substations). GICs can be calculated as follows

\begin{equation}  \label{eq:5}
 \textbf{I}=(\textbf{1}+\textbf{Y}\textbf{Z})^{-1}\textbf{J}
\end{equation}

\noindent where \textbf{1} is the unit matrix, \textbf{Y} is the network admittance matrix,
and \textbf{Z} is the earthing impedance matrix. The admittance matrix \textbf{Y} is defined by the 
resistances of the conductors of the network:

\begin{equation}  \label{eq:6}
 Y_{ij}\,=-\frac{1}{R_{ij}}, \;(i\neq j)
\end{equation}

\begin{equation}
 Y_{ij} = \Sigma_{k\neq i}\frac{1}{R_{ik}},  \;(i=j) 
\end{equation}
%  \begin{equation}  \label{eq:7}

% \end{equation}

\noindent where $R_{ij}$ is the resistance between two nodes $i$ and $j$. The column vector 
\textbf{J} has elements defined by 

\begin{equation}  \label{eq:8}
 J_i=\Sigma_{k\neq i}\frac{V_{ki}}{R_{ki}}
\end{equation}

\noindent where $V_{ki}$ is the voltage calculated from the line integral of the geoelectric field 
along the power line from point $k$ to $i$

\begin{equation}  \label{eq:9}
V_{ki}=\int_{k}^{i}\mathbf{E}\:ds
\end{equation}

Once all of these components are known, the current which flows from node $i$ to $k$ can be calculated 
as 

\begin{equation}  \label{eq:10}
 I_{ik} = \frac{V_{ij}}{R_{ij}} + \frac{(\textbf{ZI})_i - (\textbf{ZI})_j}{R_{ij}}
\end{equation}

\citet{Boteler2014} outline an amendment to the method above which allows for the modelling of GIC flow 
between levels of different voltages in a network, e.g., the flow of GICs between the 400 and 220~kV 
networks. GICs flow through both the high and low-voltage windings of a transformer, sharing a path to 
ground through the substation grounding resistance. The introduction of a new node at the neutral point 
in a transformer which connects different voltage levels of a modelled network will account for this 
flow without adding off-diagonal elements to the earthing impedance matrix \textbf{Z}, allowing for 
simple calculation. In this study, the 400kV substations were treated as wye-wye transformers using the approach of \citet{Boteler2014}. Although wye-wye transformers are not usually used to connect different voltage networks, this approach gave good agreement between the predicted and measured GIC levels at the sole Hall effect probe in Ireland. This is discussed further in Section 4.1.

It is important to note that the  Irish transmission system model used in this study is an ideal model. That is to say that each substation in the model is considered as having a single operational transformer, and that all of the transmission lines in the model are operational for all events. In reality, transformers and transmission lines are taken in and out of service regularly in the operation of a power network for maintenance or due to faults. It is difficult to model the exact operational configuration of a network with complete accuracy for historical events. As the GIC calculated in this paper are calculated using an ideal network, their values should be viewed qualitatively.

\section{Results}
In this section we outline  the performance of the different resistivity models for Ireland during the two most recent storms studied. We then examine the response of the Irish transmission system to a uniform electric field of 1~V~km\textsuperscript{-1} in different directions. Finally, we examine the effect of historical storms on the Irish power grid.

\subsection{Model Confirmation: 20-21 December 2015 and 06-07 March 2016} 
The Irish power network has had a GIC probe installed on a transformer at the Woodland (site number 4 in Figure \ref{power_network}) since September 2015. This continuously measures at 6.4 kHz, and averages these values to 1-minute bins. Since the installation of this probe, there have been few large (K$\geq$6) events which gave relatively clear GIC measurements at the probe. Two of the clearest GIC measurements were taken for the 20-21 December 2015 and 06-07 March 2016 storms, Kp 7- and 6+ events respectively. As such, they were chosen to test our GIC predictions.

The geomagnetic field in Ireland for both events was interpolated using SECS with data from Birr, Sligo, ESK, HAD, and LER. VAL data was available for both storms, but an hour of data was missing from 1200--1300 UT on 21 December 2015, so the site was omitted for that event (see Table \ref{stations_table}).

The measured GICs for both events, along with the \citet{Adam2012}, homogenous Earth and MT model calculated GICs are shown in Figures \ref{GIC1} and \ref{GIC2}. In both storm events, the \citet{Adam2012} resistivity model overestimates the measured GICs by a factor of approximately 10, despite matching the variations well. As the values given by the \citet{Adam2012} model were so large, it will not be discussed in any great detail. Qualitatively, both the homogenous and MT models give reasonable approximations for the measured GICs when the amplitude rises above the noise level ($\sim\pm$0.2~A). 

\begin{center}
\begin{figure*}
\noindent\includegraphics[width=30pc]{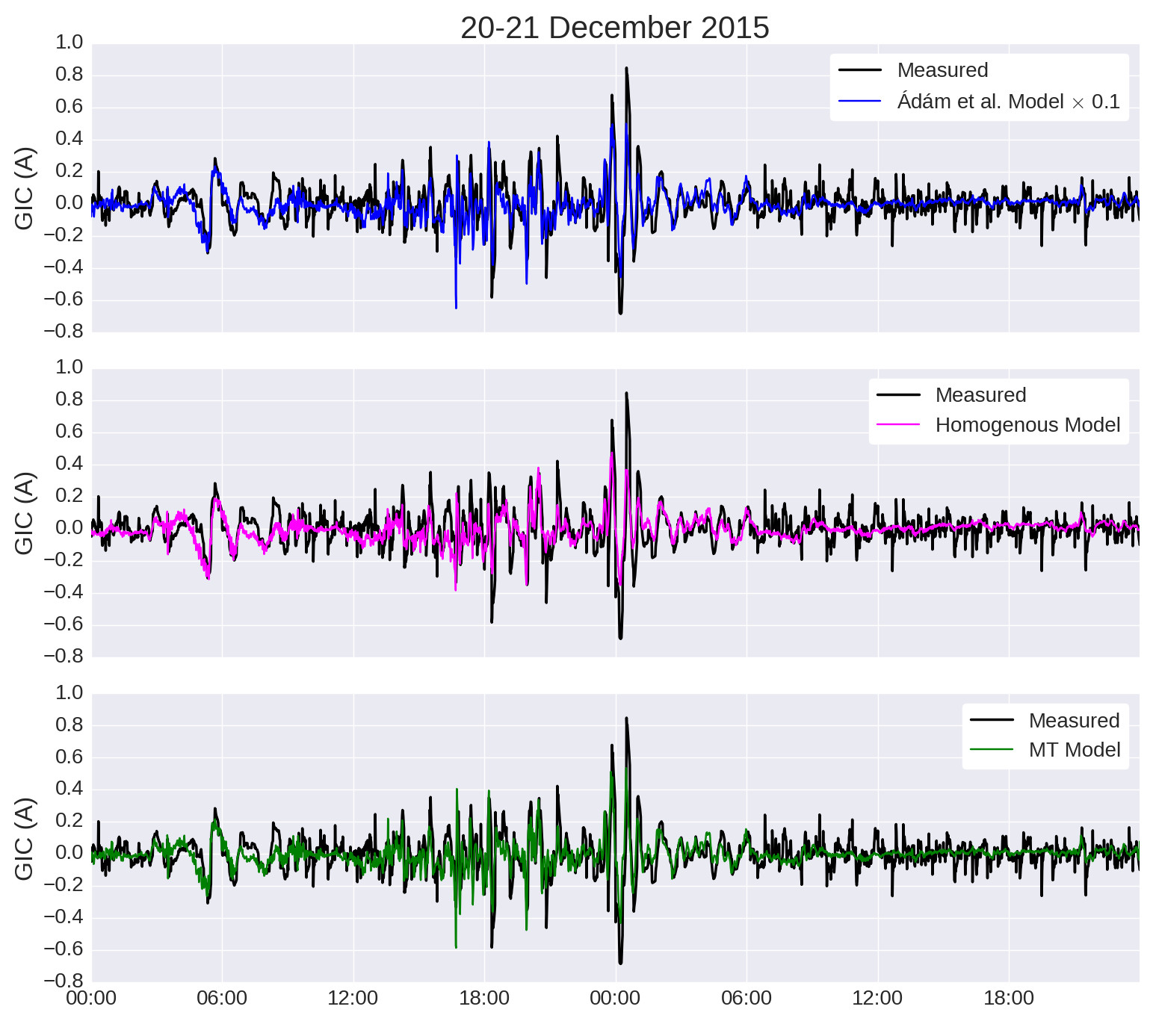}
\caption{Measured and calculated GICs at the Woodland 400 kV substation near Dublin (transformer 4 in Figure \ref{power_network}) for the 20-21 December 2015 geomagnetic event. Top: GICs calculated using the \citet{Adam2012} model. These are plotted at 10\% amplitude. Middle: GICs calculated using the 100 $\Omega$ m model. Bottom: GICs calculated using the MT derived resistivity model.}
\label{GIC1}
\end{figure*}

\begin{figure*}
\noindent\includegraphics[width=30pc]{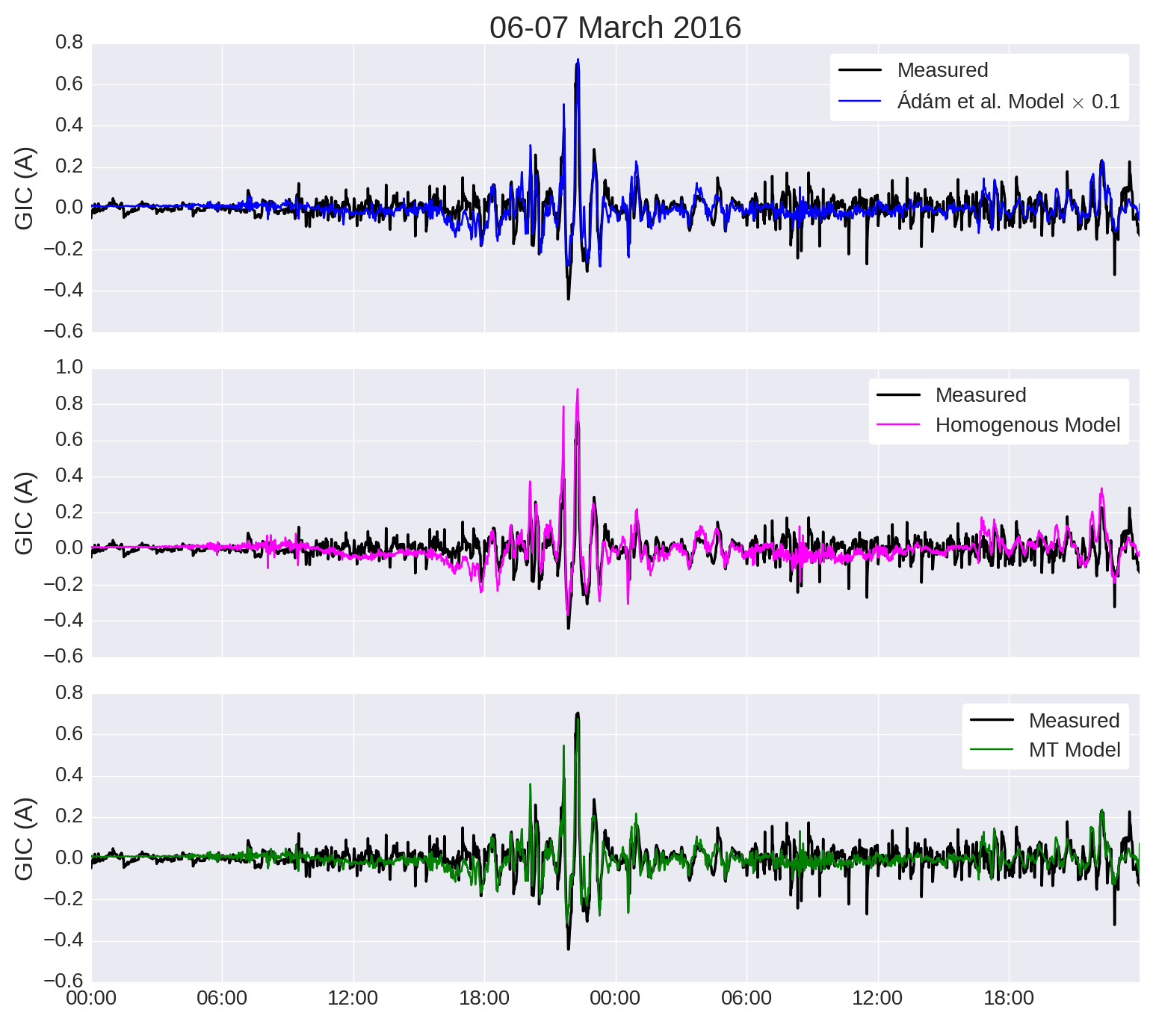}
\caption{Measured and calculated GICs at the Woodland 400 kV substation near Dublin (transformer 4 in Figure \ref{power_network}) for the 06-07 March 2016 geomagnetic event. Top: GICs calculated using the \citet{Adam2012} model. These are plotted at 10\% amplitude. Middle: GICs calculated using the 100 $\Omega$ m model. Bottom: GICs calculated using the MT derived resistivity model.}
\label{GIC2}
\end{figure*}
\end{center}
 
An important characteristic of GICs is its instantaneous magnitude \citep{Zheng2013}, as this relates with reactive power absorption in a transformer. Both models underestimate the largest GICs measured during the December 2015 event (from 23:30 UT on the 20th to 02:30 UT on the 21st), although the MT model underestimates to a lesser degree than the homogenous model. For the March 2016 storm, both models give very similar estimations. Again, the MT model fares slightly better when estimating the largest peaks, as the homogenous model overestimates these by about 30\%.

To quantify the effectiveness of the models used in this paper, a number of calculations were applied to the predicted and measured GICs. The first of these is the root mean square difference (RMSD) which is defined in Equation \ref{eq:1}. Secondly, the Pearson correlation coefficient $R$, a measure of linear correlation between two variables (where 1 is total positive correlation, 0 is no correlation, and --1 is total negative correlation) was calculated. Finally, the  \citet{Torta2014} defined performance parameter was applied to the data. The performance parameter $P$ is defined as

\begin{equation}  \label{eq:11}
P = 1 - \frac{\textrm{RMSD}_{oc}}{\sigma_{o}}
\end{equation}

\noindent where subscripts $o$ and $c$ refer to observed and calculated values, and $\sigma$ is standard deviation. A $P$ value of 1 denotes a complete match between observed and measured values. The results of these different tests are shown in Table \ref{goodness}. The MT model is shown to be marginally better than the homogenous model for all of the measurements used, although this improvement is hardly appreciable for the December 2015 event. The difference between the models is more apparent for the March 2016 storm, due to the homogenous model overestimating the largest peaks. The \citet{Adam2012} model is shown to match the variability of the two events quite well (as seen in the high Pearson correlation coefficients), but due the large amplitude differences, it scores poorly in both the RMSD and $P$ values.

\begin{table*}
\caption{Different measurements for the goodness of the homogenous earth and MT resistivity models when calculating GICs at the Woodland transformer. The measurements are; the root mean square difference (RMSD$_{oc}$), Pearson correlation coefficient (R) and \citet{Torta2014} defined performance parameter (P). The \citet{Adam2012} model calculated GICs matched the variations of the measured GICs quite well, but overestimated the amplitude by a factor of ten.}
\begin{tabular}{| c || c | c | c || c | c | c|}
    \hline
     &\multicolumn{3}{c||}{ 20-21 December 2015 } & \multicolumn{3}{|c|}{ 06-07 March 2016 }\\
     \hline \hline
    & RMSD (A) & R & P & RMSD (A) & R & P \\
    Homogenous Earth & 0.095 & 0.61 & 0.207 & 0.066 & 0.69 & 0.145 \\
    MT model & 0.093 & 0.62 & 0.214 & 0.054 & 0.73 & 0.301 \\
    \citet{Adam2012}  model & 0.786 & 0.66 & -5.6 & 0.636 & 0.73 & -7.23 \\
    \hline \hline
\end{tabular}
\label{goodness}
\end{table*}

\begin{center}
\begin{figure*}
\noindent\includegraphics[width=29pc]{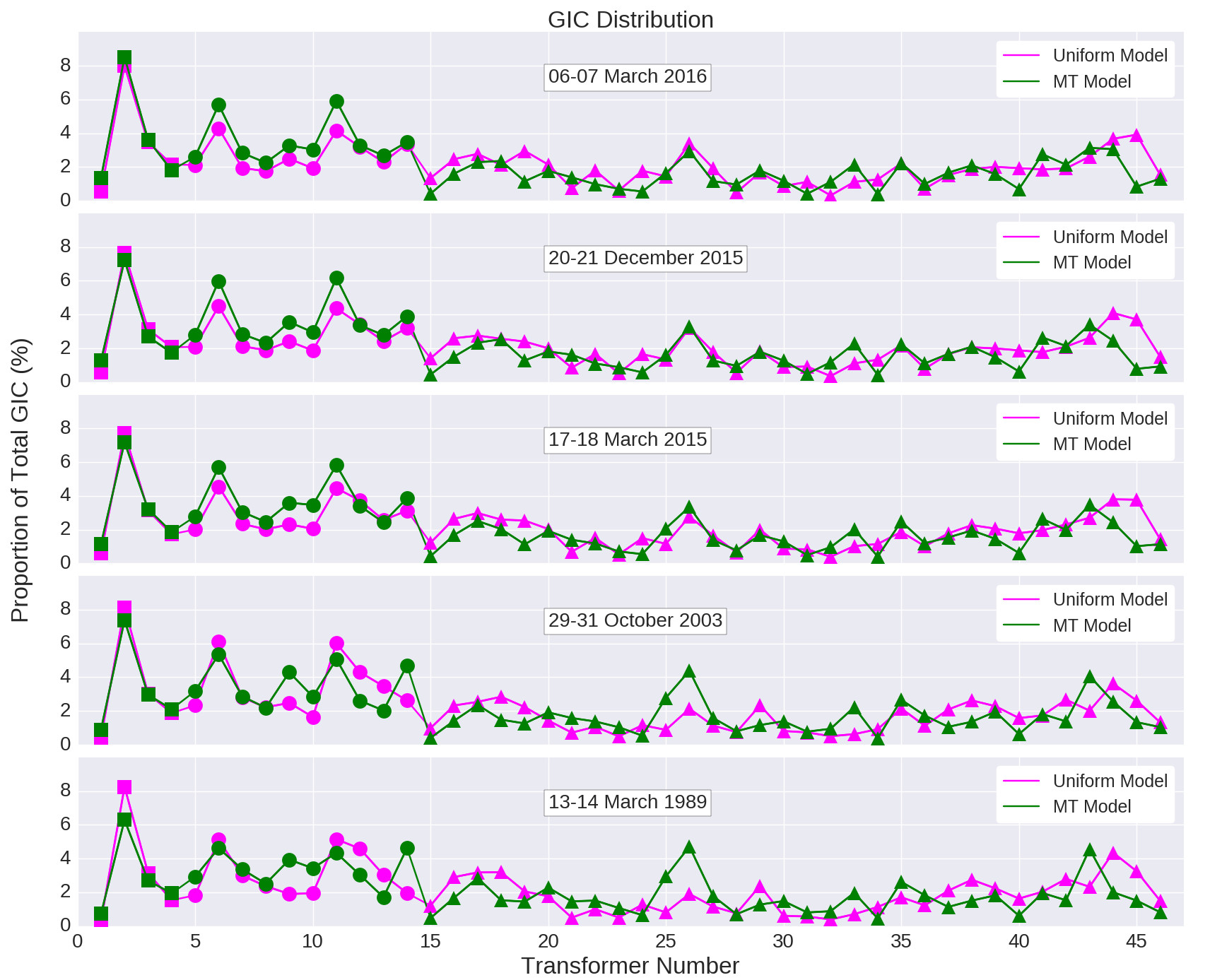}
\caption{Distribution of total GIC in the Irish power network for each of the events in this paper as calculated using the homogenous and MT resistivity models. Events are listed in reverse chronological order from top to bottom. Squares, circles and triangles represent 400~kV, 275~kV and 220~kV transformers respectively. Transformer number indicates the transformer seen in Figure \ref{power_network}.}
\label{gic_prop}
\end{figure*}
\end{center}

Both model predictions give similar GIC estimations for the rest of the network, as well as the Woodland 400~kV transformer. The distribution of GICs in the other substations in the network are given for both recent events, along with the three historical events, in Figure \ref{gic_prop}. In general, the higher voltage substations (400 and 275~kV) appear to experience proportionally more GIC than the 220~kV network for both resistivity models. This is not surprising, as higher voltage networks tend to have lower line resistance, which contributes to larger GICs. In particular, substations numbered 2, 6 and 11 each see at least 4\% of induced current in the entire network for the homogenous model, and at least 5\% for the MT model for all events. For every event, the homogenous and MT models predict that the Moneypoint substation (number 2 in Figure \ref{power_network}) will experience more GIC than any other substation. This is in keeping with the general transmission system response analysis, which will be shown in Section \ref{blank}. 

The differences between the models become more apparent in the larger storms. This can be seen in transformers numbered 14, 26 and 43 during the October 2003 and March 1989 storms. Each of the transformers experience proportionally more GIC when calculated with the MT model. This `extra' current seen in the MT model can be said to be due to the modelled geology of Ireland.

\subsection{General Transmission System Response} \label{blank}

\begin{center}
\begin{figure*}
\noindent\includegraphics[width=29pc]{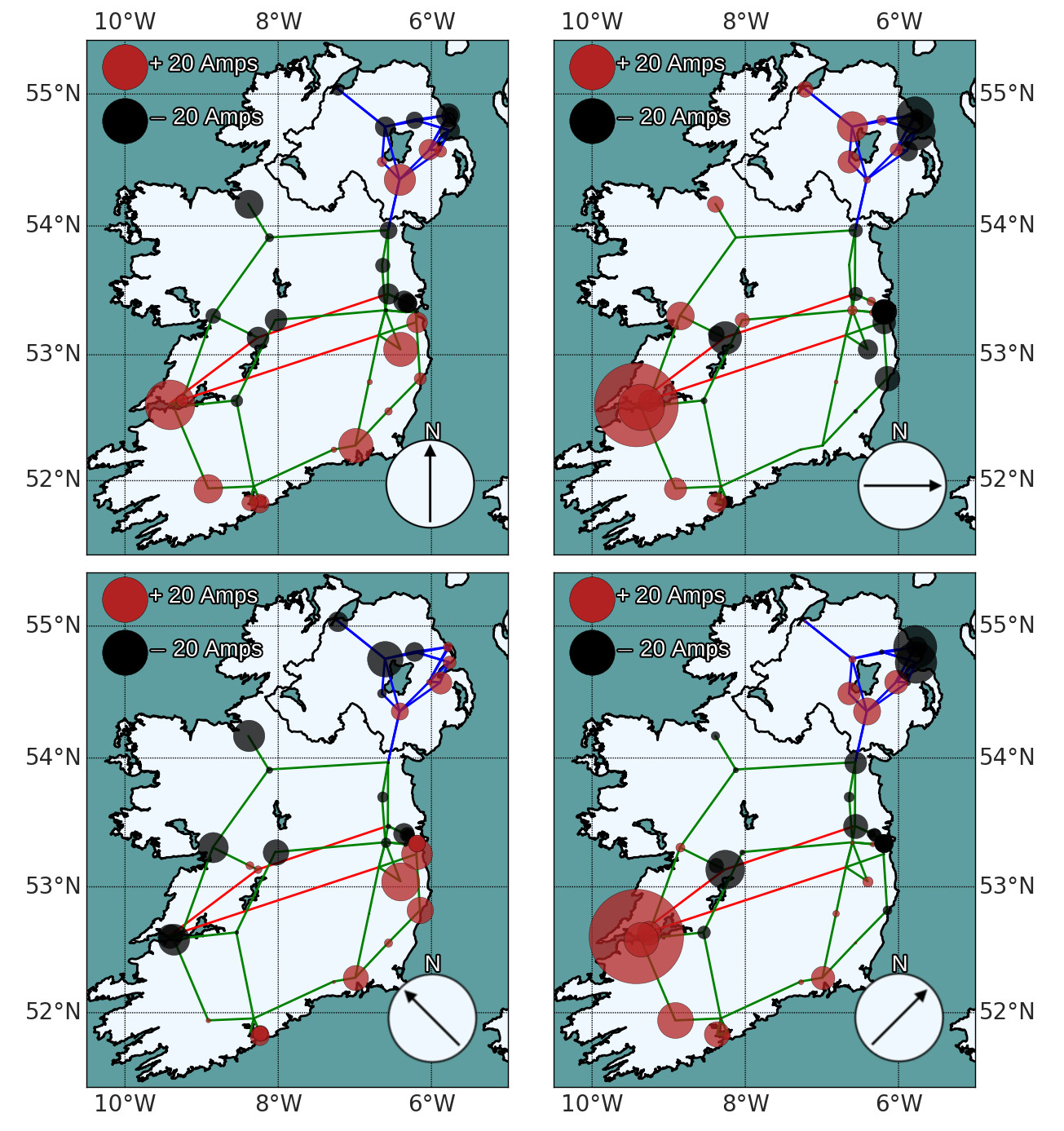}
\caption{Estimated GIC response of the Irish power grid from uniform electric fields of 1~V~km\textsuperscript{-1} directed northward (top left), eastward (top right), northwestward (bottom left) and northeastward (bottom right). Red and black circles indicate GICs which flow from and to the ground respectively. A uniform 1~V~km\textsuperscript{-1} electric field pointing northeastward generates the largest GICs in Moneypoint in western Ireland (substation number 2 in Figure \ref{power_network}) which measures 41 A.}
\label{1vkm}
\end{figure*}
\end{center}

The general susceptibility of a power network to GICs can be examined by applying a uniform
electric field of 1~V~km\textsuperscript{-1} in different directions to a region and 
subsequently calculating GICs. While it is true that electric fields across a country will 
not be uniform during an actual geomagnetic storm (due to the nature of the Earth's geomagnetic field 
and extremely variable conductivity structure), this exercise gives an indication as to which 
substations will favour GICs due solely to the orientation of the network. The results of this 
calculation for the Irish 400, 275 and 220~kV transmission system are shown in Figure \ref{1vkm}. 

The calculations show that GICs are roughly uniform in substations across the country, with the 
exception of the north-eastern 275~kV transformers and the 400~kV transformer in the west of the 
country, which experienced larger than average currents. In the case of the 400~kV transformer, this 
higher susceptibility is likely due to it being connected to the rest of the grid via the longest 
transmission lines in the country. The highest GIC calculated from a uniform electric field of 1~V~km\textsuperscript{-1} arises when the uniform field points northeastward. This is unsurprising given the NE-SW 400~kV lines which span the country. The highest GIC calculated for this case is 41~A, and was calculated for the Moneypoint 400~kV substation (numbered 2 in Figure \ref{power_network}).

\subsection{Historical Event Analyses}
Following the analysis of the different resistivity models, three well-known geomagnetic events were studied. These are the 17-18 March 2015, 29-31 October 2003 and 13-14 March 1989 storms. A summary of peak measured and calculated values for each event are given in Table \ref{summary}. More detailed breakdowns of each event are now given in reverse chronological order. Descriptions of storm variations are given in terms of the Valentia Observatory, as this is the only Irish observatory which was operating for all three historical events.

\begin{table*}
\caption{Peak measured and calculated values for each of the events studied. Both Dst and Kp values were obtained from the World Data Centre for Geomagnetism, Kyoto, (www.wdc.kugi.kyoto-u.ac.jp). The maximum $\frac{dB}{dt}$ values are from calculated SECS data. Maximum $E_h$ values were calculated using both
the homogenous 100~$\Omega$~m (superscript $H$) and MT derived (superscript $MT$) resistivity models.}

\begin{center}
    \begin{tabular}{| c || c | c | c | c | c | c | c |}
    \hline
  Date & Dst (nT) & Kp & $dB/dt$ (nT~min\textsuperscript{-1}) & $E_{h}^{H}$ (V~km\textsuperscript{-1})  & $E_{h}^{MT}$ (V~km\textsuperscript{-1}) & GIC\textsuperscript{$H$} (A) & GIC\textsuperscript{$MT$} (A) \\
  \hline  \hline
     06-07 March 2016 & -98 & 6+ & 39 & 0.08 & 0.19 & 2.4 & 3.4 \\
     20-21 December 2015 & -155 & 7- & 75 & 0.08 & 0.21 & 2.0 & 2.2 \\
     17-18 March 2015 & -223 & 8- & 128 & 0.18 & 0.51 & 2.96 & 5.8 \\ 
     29-31 October 2003 & -383 & 9o & 454 & 0.79 & 2.26 & 18.3 & 24.0 \\ 
     13-14 March 1989 & -589 & 9o & 955 & 1.46 & 3.85 & 25.8 & 23.1  \\
    \hline
    \end{tabular}
\label{summary}    
\end{center}
\end{table*}

\subsubsection{17-18 March 2015 (St. Patrick's Day Storm)}
The St. Patrick's Day storm was a Kp~8- event with a peak Dst value of -223 nT. The local K-index as measured at Birr in Ireland peaked at K7 for a period of $\sim$12 hours, but variability persisted with local K5s still measured 24 hours later. 

Of all of the events studied in this paper, the 2015 St. Patrick's Day storm had the most geomagnetic observatories operating, with two MagIE sites operating in Ireland (Birr and Leitrim) in addition to Valentia.
The storm commenced at approximately 04:45 UT with the arrival at Earth of a CME \citep{Astafyeva2015}. From 04:50 UT, there was a sharp increase of a few tens of nT in the Valentia magnetic field. The most disturbed period of the storm lasted from about 13:30 UT on 17 March 2015 until 03:00 UT the following day. Conditions continued to be disturbed until the end of the day.

\begin{center}
\begin{figure*}
\noindent\includegraphics[width=26pc]{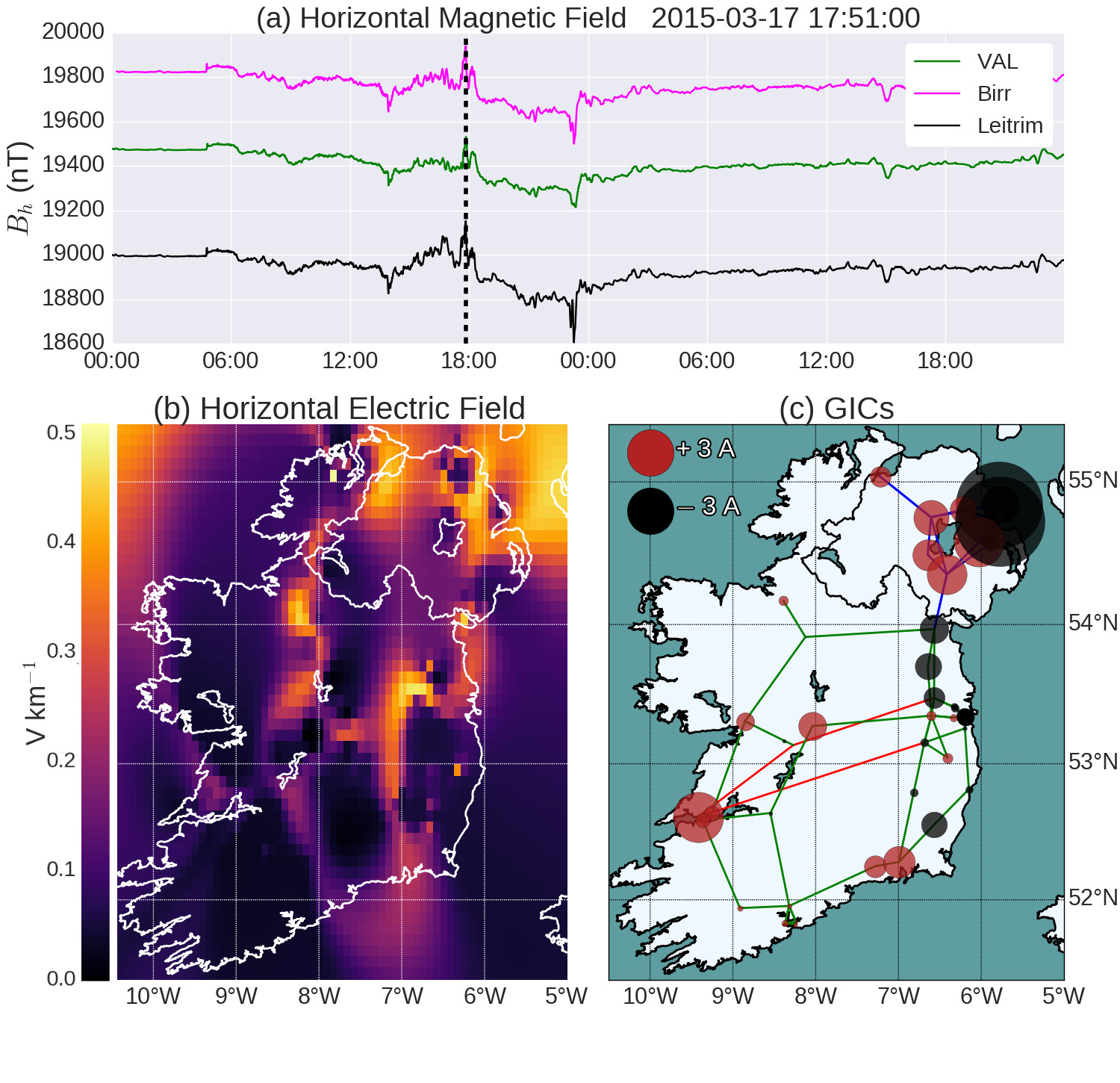}
\caption{The geomagnetic, geoelectric and GIC conditions in Ireland during the 17-18 March  2015 storm. (a) The horizontal magnetic field as measured in Valentia, Birr and Leitrim. The dashed line indicates the time displayed in the bottom of two plots. This time was chosen as it corresponds with the highest GIC calculated during the event. (b) The horizontal electric field as calculated using the MT resistivity model. Maximum electric field was calculated at 0.51~V~km\textsuperscript{-1} (c) The corresponding calculated GICs in the Irish power network. Maximum GIC were calculated at 5.8 A. A movie of this simulation is included with the paper.}
\label{MAR2015_super}
\end{figure*}
\end{center}

The SECS simulation for the event gave a maximum $dB/dt$ for the island of 128~nT~min\textsuperscript{-1}. Maximum calculated electric fields for the event were 0.18 and 0.51 V~km\textsuperscript{-1} for the homogenous and MT resistivity models respectively. Maximum GICs were calculated at 3 and 5.8~A for each of the models. A snapshot of the magnetic, electric and GIC conditions for the St. Patrick's Day storm can be seen in Figure \ref{MAR2015_super}. A movie of the simulation is given with the paper. 

\subsubsection{29-31 October 2003 (Halloween Storms)}
The 2003 Halloween Storms were a series of Kp 9 events with peak Dst values of -383~nT. Conditions were extremely disturbed from 29 October until 31 October due to a series of CMEs which erupted from the Sun in the preceding days.
At 06:30~UT on 29 October 2003, there was a drop in the magnitude of the horizontal magnetic component at Valentia of $\sim$850~nT over 30 minutes. It returned to average levels for the day, but variations of a few hundred nT over timescales less than an hour continued until 03:00~UT the following day, when conditions quietened. The second major part of the storm began at about 21:00~UT on 30 October. The horizontal geomagnetic component at Valentia was most disturbed at 00:42~UT on 31 October with a rise and fall of $\sim$900~nT measured over less than an hour. 

\begin{center}
\begin{figure*}
\noindent\includegraphics[width=26pc]{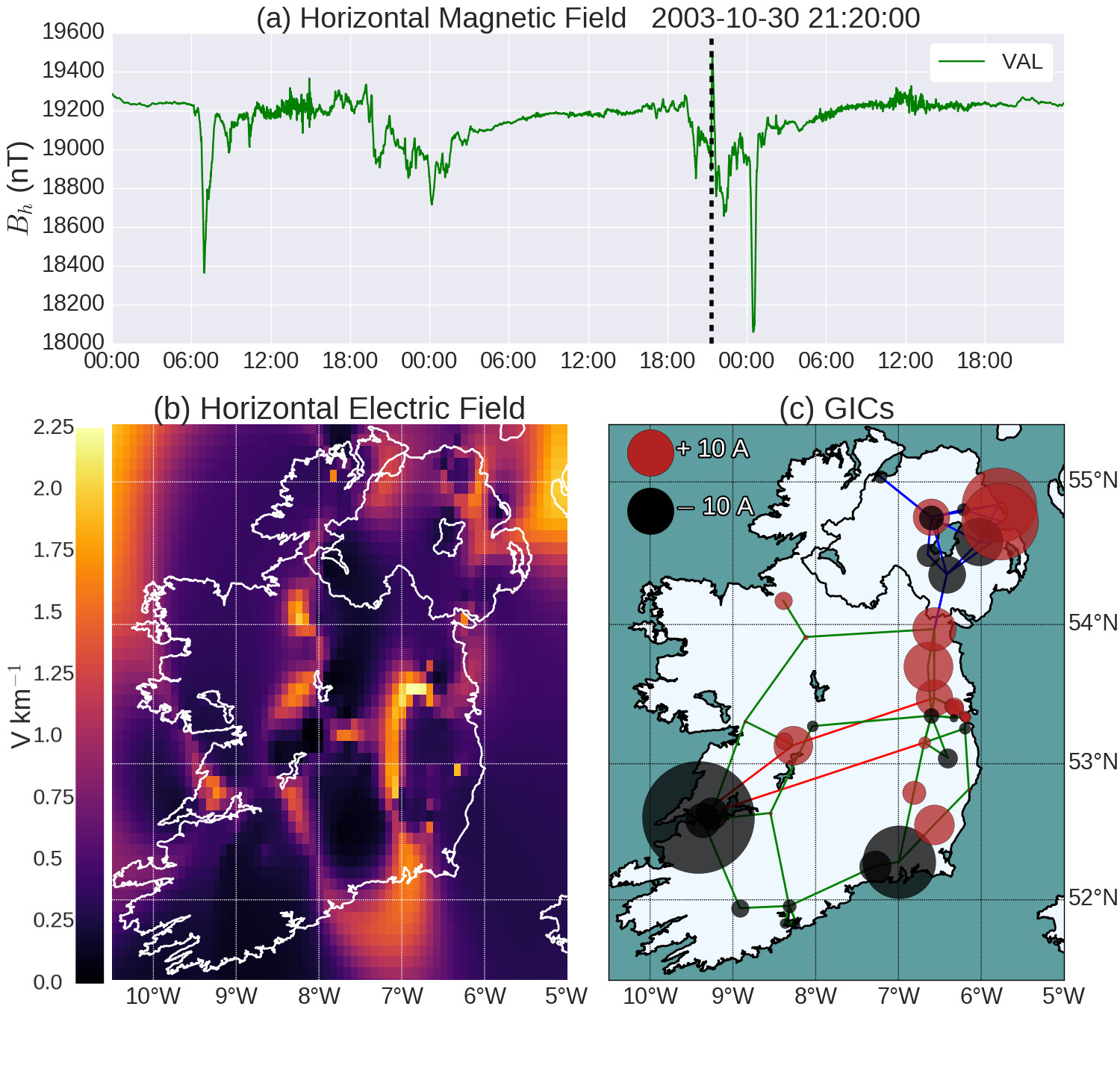}
\caption{The geomagnetic, geoelectric and GIC conditions in Ireland during the 29-31 October 2003 `Halloween' storm. (a) The horizontal magnetic field as measured in Valentia, the only Irish geomagnetic observatory operating at the time. The dashed line indicates the time displayed in the bottom of two plots. This time was chosen as it corresponds with the highest GIC calculated during the event. (b) The horizontal electric field as calculated using the MT resistivity model. Maximum electric field was calculated at 2.26~V~km\textsuperscript{-1} (c) The corresponding calculated GICs in the Irish power network. Maximum GIC were calculated at 24 A. A movie of this simulation is included with the paper.}
\label{OCT2003_super}
\end{figure*}
\end{center}

The SECS simulation for the event gave a maximum $dB/dt$ for Ireland of 454~nT~min\textsuperscript{-1}. The peak calculated E-field using the homogenous and MT resistivity models were 0.79 and 2.26~V~km\textsuperscript{-1} respectively. The two models gave peak GICs of 18.3 and 24.0~A. A snapshot of the conditions for the Halloween storms can be seen in Figure \ref{OCT2003_super}. A movie of the simulation is given with the paper.

\subsubsection{13-14 March 1989}
The March 1989 storm is the largest of the events studied, with a Kp value of 9o and a peak Dst value of -589~nT. This was the event which triggered the catastrophic loss of power in the Qu\'{e}bec power network \citep{Bolduc2002}.
Significant disturbances were measured in the horizontal magnetic field components in Valentia from approximately 07:45~UT on the morning of 13 March, with variations of about 200~nT over an hour. These continued until 20:25~UT that night, when the major part of the storm began. From 20:25~UT on 13 March until 02:00~UT the following day, variations in excess of 250~nT over timescales less than an hour were measured at Valentia.

\begin{center}
\begin{figure*}
\noindent\includegraphics[width=26pc]{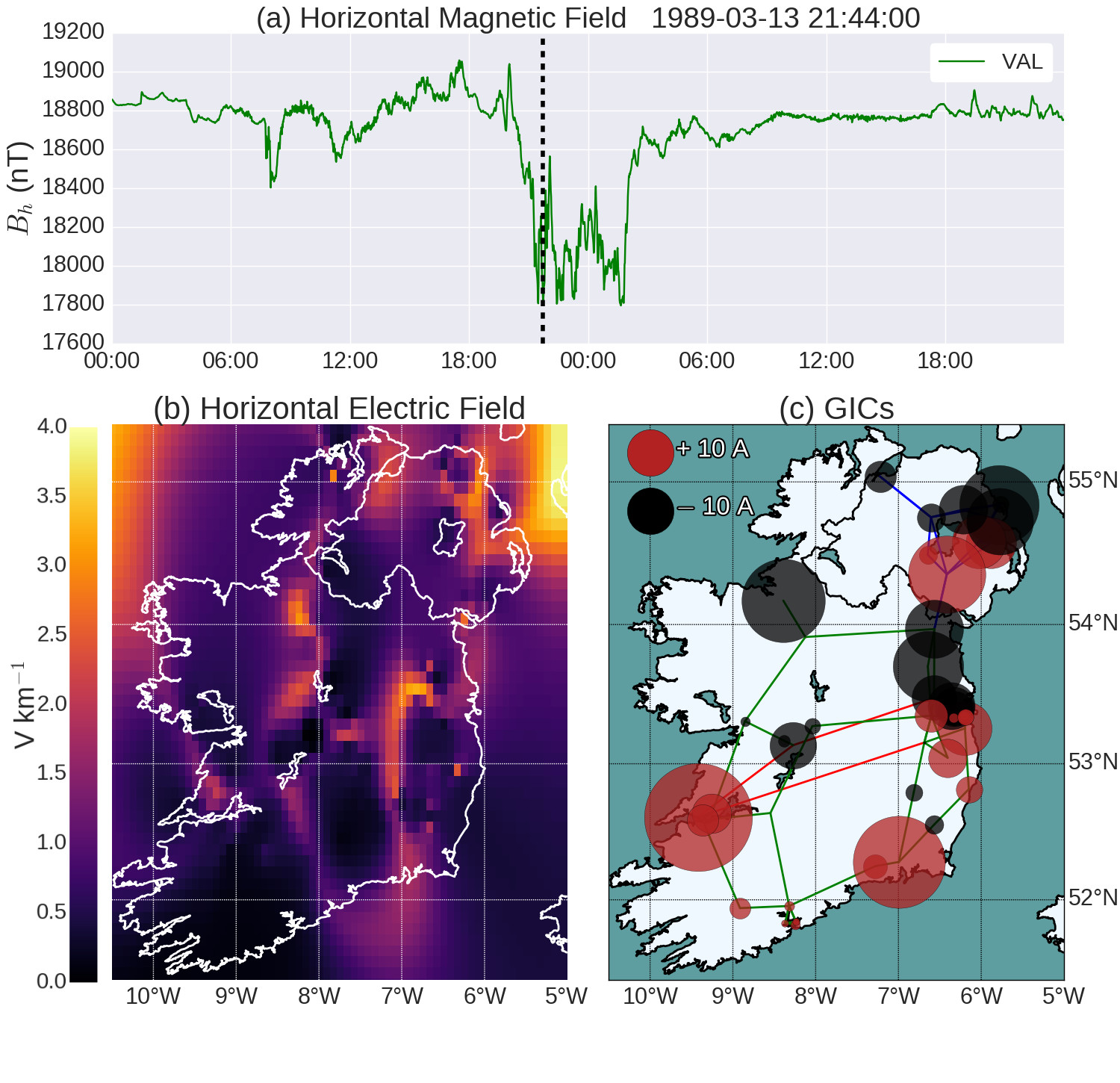}
\caption{The geomagnetic, geoelectric and GIC conditions in Ireland during the 13-14 March 1989 storm. (a) The horizontal magnetic field as measured in Valentia, the only Irish geomagnetic observatory operating at the time. The dashed line indicates the time displayed in the bottom of two plots. This time was chosen as it corresponds with the highest GIC calculated during the event. (b) The horizontal electric field as calculated using the MT resistivity model. Maximum electric field was calculated at 3.85~V~km\textsuperscript{-1} (c) The corresponding calculated GICs in the Irish power network. Maximum GIC were calculated at just over 23 A. A movie of this simulation is included with the paper.}
\label{MAR1989_super}
\end{figure*}
\end{center}

The SECS simulation for the event gave a maximum $dB/dt$ of $\sim$955~nT~min\textsuperscript{-1}. The peak calculated E-field using the homogenous and MT resistivity models were 1.46 and 3.85~V~km\textsuperscript{-1}, with corresponding peak GICs of 25.8 and 23.1~A. When using the MT resistivity model, the 1989 storm gave larger simulated electric fields than the October 2003 storms, despite giving a smaller peak GIC value. One would generally expect larger GICs with larger electric fields. In this instance, the 1989 storm was calculated to give larger GICs in the Irish network as a whole when compared to the 2003 storms, with 35 of the 46 substations experiencing larger peak GICs during the March 1989 event. A snapshot of the conditions for the 1989 storm can be seen in Figure \ref{MAR1989_super}. A movie of the simulation is given with the paper.

\section{Discussion}
The combination of Ireland's mid-latitude location and small area (approximately 300$\times$500~km) both act to limit the magnitude of simulated GICs in the Irish power network. The calculated GICs for the three historical events certainly give lower GIC amplitudes than can be found in larger countries, and those at more northerly latitudes (e.g., \citet{Wik2008, Myllys2014, Torta2014}). For example, during the October 2003 storms, Scotland's power grid (which was at a similar geomagnetic latitude as Ireland; 58.9$^\circ$ N for Scotland, 56.8$^\circ$ N for Ireland) saw GICs of 42~A \citep{Thomson2005}, whereas predicted peak GICs in Ireland for the same event were 24~A.

Another illustration of how Ireland's small size affects GIC calculations is to compare it with both the Norwegian and Spanish networks. From Equation \ref{eq:9}, the voltage between two nodes in a network is calculated by integrating the electric field along the path of the line, with longer lines allowing larger voltages to drive GICs. In territories with longer lines, there tends to be fewer transformers to limit GIC flow. When a uniform 1~V~km\textsuperscript{-1} electric field is applied to Ireland (therefore ignoring the influence of geomagnetic latitude on the geoelectric field), the maximum GICs calculated are 41~A. In Norway, with its transmission lines covering a much larger area (roughly 500 $\times$ 1000~km), that value is 151~A \citep{Myllys2014}. According to \citet{Myllys2014}, the total length of 400 and 300~kV lines in Norway is 7116~km. The total length of 400, 275 and 220~kV lines in Ireland amounts to under half that at 3176~km.  Similarly, a uniform electric field applied to Spain (approximately 1000 $\times$ 1000~km) generates GICs up to 153~A \citep{Torta2014}. It is clear that much larger electric fields are needed in order to generate comparable GICs in the Irish network.

Although they were two of the largest geomagnetic storms in the last half century, neither the March 1989 nor October 2003 storms produced GICs in our simulations which are likely to have been large enough to cause a catastrophic failure in transformers in Ireland. Indeed, no transformer failures were reported immediately after either event. Despite the low GIC levels, a number of substations were calculated to have multiple periods with GICs of around 15~A (particularly substations numbered 2, 6 and 11 in Figure \ref{power_network}). It is possible that these GIC levels presented opportunities for transformer heating to occur.  A detailed statistical analysis of transformer failures in Ireland would be useful to quantify the impact of geomagnetic storms in Ireland.

The event analyses, when coupled with the general transmission system response calculations, allow for the identification of substations in Ireland which may be particularly at risk from geomagnetic storms. From the application of the uniform electric field (in Section \ref{blank}), the substation at  the western end of Ireland's only 400~kV lines experiences the largest GICs. In the case of this substation (Moneypoint, number 2 in Figure \ref{power_network}), the 400~kV lines connected to it are oriented roughly perpendicular to the magnetic north-south axis. This leaves the lines particularly sensitive to changes in the magnetic field along this axis. Substations numbered 6 and 11 also saw proportionally more GICs for each event and resistivity model, when compared with the rest of the network.

The installation of the GIC probe at the Woodland substation in the east has allowed for the verification of our GIC calculations using both the homogenous and MT derived resistivity models. Unfortunately, the location and timing of the installation of the Hall probe are not optimal for our purposes. Firstly, from the general transmission system response analysis, Woodland would not be considered a high-risk substation: from the orientation of the power grid alone, only a small proportion of GIC would be expected to flow through the substation for all uniform electric field directions. Couple this with small sample of significant geomagnetic storms since its installation date (September 2015), and the result is usable GIC data which is only four times the noise level of the Hall effect probe. This can be seen in Figures \ref{GIC1} and \ref{GIC2}, with the noise level of $\pm$0.2~A.

For the homogenous resistivity model, a half-space of 100~$\Omega$~m was chosen, as it best fit our GIC observations. Changing the value of this resistivity to 200~$\Omega$~m would change the calculated GIC at Woodland by a small amount (0.2~A for the December 2015 event), but will change the calculated GIC at a substation such as Moneypoint by a much larger amount (0.83~A for the same event). As such, fitting to a small noisy signal in Woodland for two minor events has larger implications for the rest of the Irish power grid, and caution should be used when drawing conclusions as to the effectiveness of resistivity models for the whole of the network.

That said, for our limited sample of measured GICs, the MT derived resistivity model was seen to perform marginally better than the homogenous model at replicating measured data, particularly during the March 2016 event. Functionally, the two models give quite similar GIC signals for the two events. It is a well-known side-effect of Equation \ref{eq:9} that integrating the electric field along a transmission line effectively `smooths' the electric field between nodes, meaning that a high resolution conductivity model may not be drastically more accurate than a more simple model, despite having more accurate surface electric fields \citep{Viljanen1994}. This can be seen clearly in Table \ref{summary}. Both models show similar peak GICs, despite having quite different peak electric fields.

The MT resistivity model used in this paper is itself not particularly sophisticated, despite being derived from multiple real world MT measurements. It is made up from a simple interpolation of data from MT sites. This means that the interpolation may not be accurate for regions not bounded by MT data. A more realistic approach might be to constrain the interpolation with known geological data, such as was done in \citet{Beamish2013}.

The plane-wave method used to calculate electric fields in this study has its limitations. It does not account for spatial resistivity effects. The electric field is calculated at each 10$\times$10~km square independently which means that important phenomena such as the coastal effect are neglected when calculating the electric field. For a small island such as Ireland, this effect may be quite important in affecting electric fields along or near the coastlines. This is particularly important given the orientation of the Irish power grid, which follows the coastlines. Other methods such as the thin-sheet approximation take into account the spatial conductivity of a region, and can be used for GIC studies \citep{Thomson2005}.

The model network used in this study is a first approximation of the Irish power grid. Each substation in the model is assumed to have a single transformer in operation. It also assumes that all of the transmission lines were in operation for each of the events. In reality, substations often have multiple transformers operating, and transmission lines are frequently taken down for maintenance. Future work will take into account a greater level of detail in our representation of the Irish power network, with different transformer types being modelled appropriately, and correct system configuration for particular historical events.

\section{Conclusion}
This study is the first to simulate GICs in the Irish power transmission system for multiple severe geomagnetic storms. Electric fields throughout Ireland were estimated using the plane-wave method coupled with an MT derived multi-layered resistivity model, as well as with a homogenous Earth resistivity model. GICs were replicated for two recent K6+ and K7- events in a transformer in the east of the country. While both resistivity models performed well in replicating the measured GIC, the MT derived model was seen to perform marginally better than its homogenous counterpart.

Using the MT and homogenous resistivity models, three historical storms were simulated. These were 
the 17-18 March 2015, 29-31 October 2003 and 13-14 March 1989 storms. Of all of the events studied, the 30-31 October 2003 and 13-14 March 1989 storms  each gave GIC values which may have contributed to transformer heating. Peak GICs for these events were calculated at 18.3 and 25.8 A respectively using the homogenous Earth model, while the MT model gave peak values of 24.0 and 23.1 A for the two storms respectively. 

Using the multiple storm analyses along with a general transmission system, a number of transformers were identified as being most likely to experience larger GICs in Ireland. These are the 400~kV Moneypoint, 275~kV Ballylumford and 275~kV Kilroot substations (numbered 2, 6 and 11 in Figure \ref{power_network}).

While this study gives an indication as to the level of GICs that can be expected for Kp8 and Kp9 storms, a statistical analysis of Ireland's historical geomagnetic field is required to quantify the GIC risk for Ireland over large time scales.

Future geomagnetic storms will now be measured at multiple sites in Ireland, including the Valentia, Birr, Sligo and Armagh magnetic observatories. In addition, the GIC probe at the Woodland substation will continue to directly measure GICs in Ireland, increasing the sample size of storms to study in Ireland with time.

%%% End of body of article:

%%%%%%%%%%%%%%%%%%%%%%%%%%%%%%%%
%% Optional Appendix goes here
%
% \appendix resets counters and redefines section heads
% but doesn't print anything.
% After typing \appendix
%
%\section{Here Is Appendix Title}
% will show
% Appendix A: Here Is Appendix Title
%
%%%%%%%%%%%%%%%%%%%%%%%%%%%%%%%%%%%%%%%%%%%%%%%%%%%%%%%%%%%%%%%%
%
% Optional Glossary or Notation section, goes here
%
%%%%%%%%%%%%%%
% Glossary is only allowed in Reviews of Geophysics
% \section*{Glossary}
% \paragraph{Term}
% Term Definition here
%
%%%%%%%%%%%%%%
% Notation -- End each entry with a period.
% \begin{notation}
% Term & definition.\\
% Second term & second definition.\\
% \end{notation}
%%%%%%%%%%%%%%%%%%%%%%%%%%%%%%%%%%%%%%%%%%%%%%%%%%%%%%%%%%%%%%%%
%
%  ACKNOWLEDGMENTS

\begin{acknowledgments}

This research was funded by the Irish Research Council's Enterprise Partnership Scheme between Trinity 
College Dublin and Eirgrid Plc. The results presented in this paper rely on data collected at magnetic 
observatories. We thank the national institutes that support them and INTERMAGNET for promoting high 
standards of magnetic observatory practice (www.intermagnet.org). For the ground magnetometer data for 
the other sites during the March 1989 storm, we gratefully acknowledge the World Data Centre for 
Geomagnetism in Edinburgh. We acknowledge the Coillte Ltd. for their help and cooperation in choosing a 
long-term quiet site to measure both electric and magnetic fields in Ireland. We also acknowledge 
Armagh Observatory for hosting a magnetometer which contributed to this work, and Met \'{E}ireann for 
Valentia Observatory magnetic measurements. Data from the GIC probe in Woodland for the December 2015 and March 2016 events can be requested from the corresponding author. Magnetometer data from Valentia Observatory for the 1989 storm can be requested from Met \'{E}ireann. Magnetometer data from MagIE observatories can be requested from the corresponding author.

\end{acknowledgments}

\end{article}

%\begin{figure}
%    \centering
%    \includegraphics[width=0.8\textwidth,natwidth=610,natheight=642]{mag_sites.pdf}
%\end{figure}

% 

% 

% 

% 

% 

% 
% 

% 

% 

% 

%
%% Enter Figures and Tables here:
%
% DO NOT USE \psfrag or \subfigure commands.
%
% Figure captions go below the figure.
% Table titles go above tables; all other caption information
%  should be placed in footnotes below the table.
%
%----------------
% EXAMPLE FIGURE
%
%\begin{figure}
%\includegraphics[width = 20pc]{mag_sites.png}
%\caption{Caption text here}
%\label{figure_label}
%\end{figure}
%
% ---------------
% EXAMPLE TABLE
%

% 

% See below for how to make sideways figures or tables.

\end{document}